\documentclass[aps,twocolumn,prl,superscriptaddress,amsmath,showpacs,tightenlines,pdflatex,longbibliography]{revtex4-1}
\usepackage{xcolor}
\usepackage{amsmath}

\usepackage{graphicx,times}
\usepackage{amstext}
\usepackage{amsmath}            
\usepackage{amssymb}            
\usepackage{graphicx}           
\usepackage{latexsym}
\usepackage{bm}

\usepackage{etoolbox}
\usepackage{breqn}

\makeatletter
\let\cat@comma@active\@empty
\makeatother

\newcommand{\revision}[1]{{#1}}

\newcommand{\beq}{\begin{eqnarray}}
\newcommand{\eeq}{\end{eqnarray}}

\newcommand{\eq}[1]{Eq.~(\ref{#1})}

\def\<{\langle}
\def\>{\rangle}

\def \info#1{}

\begin{document}

\title{Quantifying Non-Markovianity with Temporal Steering}

\author{Shin-Liang Chen}
\affiliation{Department of Physics and National Center for
Theoretical Sciences, National Cheng-Kung University, Tainan 701,
Taiwan}

\author{Neill Lambert}
\affiliation{CEMS, RIKEN, 351-0198 Wako-shi, Japan}

\author{Che-Ming Li}
\affiliation{Department of Engineering Science and Supercomputing
Research Center, National Cheng-Kung University, Tainan City 701,
Taiwan}

\author{Adam Miranowicz}
\affiliation{CEMS, RIKEN, 351-0198 Wako-shi, Japan}
\affiliation{Faculty of Physics, Adam Mickiewicz University,
61-614 Pozna\'n, Poland}

\author{Yueh-Nan Chen}
\email{yuehnan@mail.ncku.edu.tw} \affiliation{Department of
Physics and National Center for Theoretical Sciences, National
Cheng-Kung University, Tainan 701, Taiwan} \affiliation{CEMS,
RIKEN, 351-0198 Wako-shi, Japan}

\author{Franco Nori}
\affiliation{CEMS, RIKEN, 351-0198 Wako-shi, Japan}
\affiliation{Department of Physics, The University of Michigan,
Ann Arbor, Michigan 48109-1040, USA}

\begin{abstract}
Einstein-Podolsky-Rosen (EPR) steering is a type of quantum
correlation which allows one to remotely prepare, or steer, the
state of a distant quantum system. While EPR steering can be
thought of as a purely spatial correlation there does exist a
temporal analogue, in the form of single-system temporal steering.
However, a precise quantification of such temporal steering has
been lacking. Here we show that it can be measured, via
semidefinite programming, with a {\em temporal steerable weight},
in direct analogy to the recently proposed EPR steerable weight.
We find a useful property of the temporal steerable weight in that it
is a non-increasing function under completely-positive
trace-preserving maps and can be used to define a
 sufficient and practical measure of strong non-Markovianity.
\end{abstract}



\maketitle

Quantum entanglement, Einstein-Podolsky-Rosen (EPR) steering, and
Bell non-locality are three of the most intriguing phenomena in
quantum physics and, in varying degrees, are thought to act as
resources;  fuel that powers a range of quantum technologies.
Entanglement~\cite{QE_Horodecki,Wiseman07,Jones07} comes in hand-in-hand
with the complexity of quantum systems, and may be behind the
potential speed-up of quantum computation. Bell non-locality and
EPR steering are thought to be the driving power of quantum
cryptography, and have both been recast in that language.  For
example, in a quantum key distribution scenario, two parties wish
to generate a secret key using shared quantum states as a
resource. If one party (Bob) trusts his own experimental apparatus
but not that of the other party (Alice), a violation of a steering
inequality~\cite{Cavalcanti09, Wiseman07,Jones07} can be used to certify
that true quantum correlations exist between their shared states.
In stricter terms, such a test proves to Bob that the correlations
he observes between his measurement results and Alice's cannot be
described by a local hidden state model;  his state is truly being
influenced by Alice's measurements in a non-local manner. As with
entanglement, one quantify the amount of steering that is possible
with a given shared state via a range of possible
measures~\cite{Sania,Kogias15,He15,Piani15}.  Very recently, a powerful
example of such a measure, the steerable weight, was proposed by
Skrzypczyk \textit{et al}.~\cite{Paul14,Pusey13}.

In EPR steering the notion of non-locality, via space-like
separations between parties, plays an important role. If we relax
this constraint, and consider time-like separation of measurement
events, can the concept of steering still be used in a meaningful
way?  We can find inspiration in the fact that there do already
exist other types of non-trivial temporal quantum correlations
complementary to both Bell non-locality and entanglement. For the
former, one of the most well-known examples is the Leggett-Garg
(LG) inequality~\cite{LG85}, which can be used to test the
assumption of ``macroscopic realism'', in contrast to the
non-local realism tested by Bell's Inequality, and for which
experimental violations have been observed in a large range of
systems~\cite{Palacios-Laloy10,Knee2012,Emary14}. For the latter,
motivated by the Choi-Jamiolkowski (CJ)
isomorphism~\cite{Jamiolkowski72}, which equates the correlations
in a bi-partite quantum system with two-time correlations of a
single quantum system, the notion of temporal entanglement has
been proposed in various
forms~\cite{Brukner04,Fritz10,Olson11,Olson12,Sabin12,Megidish13,Fitzsimons13}.
Returning to steering, the concept of temporal steering, and a
temporal steering inequality, was recently introduced by Chen {\em
et al.}~\cite{YN14}. Also inspired by the CJ isomorphism, they
showed that, even without the assumption of non-locality, the
concept of one party not trusting the earlier measurements made by
another party delineates between certain classical and quantum
correlations. Not only does this have direct practical
applications in verifying a quantum channel for quantum key
distribution (QKD),  it was recently shown that temporal steering,
like EPR steering~\cite{Brunner,Uola}, is intimately linked to the
concepts of realism and joint
measurability~\cite{CM14,Pusey2015,Piani,Karthik}.

Still lacking however is a measure to quantify these ``temporal
steering'' quantum correlations. Here, in analogy to the EPR
steerable weight~\cite{Paul14,Pusey13}, we define the
temporal-steerable-weight (temporal-SW) as a measure of temporal
steering. We prove that the temporal-SW is
non-increasing under a completely-positive trace-preserving (CPT) map
and can be used to define a sufficient but not necessary measure of non-Markovianity. In
the same way that the spatial steerable weight can be considered a
measure of strong entanglement, since not every entangled state is
steerable, we define the {\em temporal}-SW as a measure of {\em
strong} non-Markovianity because it vanishes for weak
non-Markovian process. (This is also in analogy to, e.g., the phenomenon of \emph{strong
non-classicality}, which can be detected and quantified by a weaker criterion of
non-classicality~\cite{Dodonov02book,Arvind97}). We show this by comparing the
non-Markovianity measured by the temporal-SW to an existing
entanglement-based measure~\cite{Rivas10}, and find that it is, as expected,
less-sensitive.  However, the temporal-SW is, in principle, easier
to implement experimentally, as it does not require the use of an
ancilla, nor full process tomography. These results, together with
with a few illustrative examples discussed mainly in the
Supplementary Material~\cite{Supplement}, suggest that temporal
steering can serve as a unique and useful quantum resource.

\begin{figure}
\includegraphics[width=1\columnwidth]{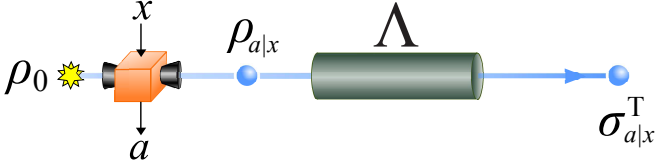}

\caption{(Color online) Schematic diagram of temporal steering. In
the beginning, Alice performs the measurement
$F_{a|x}=M_{a|x}^{\dagger}M_{a|x}$ on an initial state
$\protect\rho_{0}$. Then, $\rho _{0}$ is mapped to $\protect\rho
_{a|x}$ and sent into a quantum channel $\Lambda $. Finally, Bob
receives the assemblage $\{\sigma _{a|x}^{\text{T}}\}$ at time
$t$.}
\end{figure}
\emph{Temporal steerable weight.---} Now we introduce the concept
of a temporal steerable weight in analogy to the spatial steerable
weight introduced recently by Skrzypczyk \textit{et
al}.~\cite{Paul14,Pusey13}. In the standard (spatial) EPR steering
scenario, Alice performs a POVM (positive-operator valued measure)
measurement $F_{a|x}=M_{a|x}^{\dagger}M_{a|x}$, $\sum_a F_{a|x}
=\openone$, on a state $\rho_{AB}$ shared with Bob and creates the
assemblage $\{\sigma _{a|x}\}$, where $a$ is the measurement
result and $x$ is the basis of the measurement. In defining a
steering inequality, or a steerable weight, one assumes that Bob
does not trust Alice, nor her experimental apparatus, and wishes
to distinguish between true manipulation of his local state via
quantum correlations and correlations that cannot be distinguished
from some classical theory, typically a local hidden state model.
In temporal steering, we also let Alice perform a POVM measurement
$F_{a|x}=M_{a|x}^{\dagger}M_{a|x}$ but on a single system in an
initial state $\rho _{0}$ at time $t=0$. After the measurement,
the initial state is mapped to $\rho _{a|x}$ (see Fig.~1):
\begin{equation}
\rho _{0}\mapsto \rho _{a|x}=\frac{M_{a|x}\;\rho _{0}\;M_{a|x}^{\dagger }}{p(a|x)},
\label{mapping}
\end{equation}
with the probability $p(a|x)=\text{tr}(M_{a|x}\rho
_{0}M_{a|x}^{\dagger})$.  After this initial measurement, the
state $\rho _{a|x}$ is sent into a quantum channel $\Lambda $ for
a time $t$. At time $t$, Bob receives the system and performs
quantum tomography to obtain the state $\sigma_{a|x}$, i.e.,
$\Lambda (\rho _{a|x})=\sigma_{a|x}$. To mimic the unnormalized
assemblage~\cite{Paul14,Pusey13} in standard EPR-steering, we
define the unnormalized states in temporal steering
\begin{equation}
\sigma _{a|x}^{\text{T}}\equiv p(a|x)\;\sigma_{a|x},
\label{subnormalized}
\end{equation}
where the superscript $\text{T}$ reminds one that the assemblage $\{\sigma _{a|x}^{\text{T}}\}$ is for temporal steering.

However, the quantum channel may be noisy, obliterating the
influence of Alice's measurement choice, or Alice's measurement
results could have been fabricated via classical strategies. In
these cases, $\sigma _{a|x}^{\text{T}}$ may include, or be
entirely described by, an unsteerable assemblage which we define
as
\begin{equation} 
\sigma _{a|x}^{\text{T,US}} =\sum_{\lambda
}\;P(\lambda)\;P(a_{|x}|\lambda)
\;\sigma_{\lambda }
\label{LHS}
\end{equation}
where $\sum_{\lambda}P(\lambda)=1$.  We have written the result
$a$, conditional on the basis $x$, with a subscript notation
$a_{|x}\equiv a|x$.  In the EPR setting $\lambda$ represents a
local hidden variable which determines the possible correlations
between Alice's  and Bob's measurement results from a source which
obeys classical realism. As in that case, when Alice reveals her measurement
results, Bob can update his knowledge of his state, as indicated by two equal
forms (by applying the chain rule),
$\sum_{\lambda
}P(\lambda)P(a_{|x}|\lambda)
\sigma_{\lambda }  
=\sum_{\lambda }P(a|x)P(\lambda|a_{|x}) \sigma_{\lambda }.$
Then
the unsteerable states are those states which obey the classical
(realism) chain rule for Alice's joint measurement results, as
shown in a recent work on steering witnesses~\cite{CM14}.
\revision{No matter what happens during the transmission, Bob's task is to check whether the assemblage he receives can be written in the hidden-state form [Eq. (3)] or not. If he can, this means the state Bob receives is independent of the basis $x$ Alice chooses to measure in. As mentioned above, this may be because the quantum channel is too noisy, such that the influence of Alice's measurements is no longer discernable, or Alice's measurement results could have been fabricated via classical strategies. On the other hand, if the assemblage Bob receives cannot be written in the form of Eq.~(\ref{LHS}), he is convinced that Alice has influenced his state by her choice of measurement. In this case, we call the assemblage Bob receives ``temporally steerable" and is symbolized as $\{\sigma
_{a|x}^{\text{T,S}}\}$. }

To determine the steerable weight, one
considers the overlap between the state Bob receives and the
unsteerable assemblage, such that his state can be written as a
mixture
\begin{equation}
\sigma _{a|x}^{\text{T}}=\mu \: \sigma _{a|x}^{\text{T,US}}+(1-\mu )\sigma _{a|x}^{%
\text{T,S}}.
\end{equation}%
To quantify the ``steerability in time'' for a given assemblage
$\{\sigma _{a|x}^{\text{T}}\}$, one has to maximize $\mu $, i.e.,
maximize the proportion of $\sigma _{a|x}^{\text{T,US}}$. Then,
the ``temporal steerable weight'' can be defined as
$\text{TSW}=1-\mu ^{\ast }$, in which $\mu ^{\ast }$ is the
maximum of $\mu $ and can be obtained from semidefinite
programming~\cite{SDP,Paul14,Pusey13}:
\begin{equation}
\begin{aligned} \mu^{\ast} =~ \text{max } ~~&\text{tr}\sum_{\lambda}\tilde{\sigma}_{\lambda} \\
\text{subject to }~~ &\sigma_{a|x}^{\text{T}} -
\sum_{\lambda}D_{\lambda}(a|x) \tilde{\sigma}_{\lambda}\geq 0 ~~&&\forall
~a,x\\ &\tilde{\sigma}_{\lambda}\geq 0 &&\forall ~\lambda, \end{aligned}  \label{SDP}
\end{equation}
where $\tilde{\sigma}_\lambda=\mu\sigma_\lambda$, and $D_{\lambda}(a|x)$
are the extremal deterministic values~\cite{Paul14}  of the
conditional probability distributions $P(a_{|x}|\lambda)$.
\revision{Equation~(\ref{SDP}), which is formulated as a semi-definite program (SDP), can be numerically implemented in  various  convex optimization packages, e.g., Refs.~\cite{Grant08,Andersen14}.}

So far, the formalism is parallel to the standard EPR steerable
weight~\cite{Paul14}. The primary difference is that $\{{\sigma
}_{a|x}\}$ in Ref.~\cite{Paul14} is created through the
entanglement between Alice and Bob. Here, $\{\sigma
_{a|x}^{\text{T}}\}$ is created through Alice's measurement and
the influence of the quantum channel $\Lambda $.  In the
Supplementary Material~\cite{Supplement}, we give an explicit
pedagogical example of how to evaluate the temporal steerable
weight.

\emph{Measure of non-Markovianity.---} Now we apply the introduced
temporal steering weight as a measure of non-Markovianity.
Non-Markovianity is a term used to describe the situation when an
environment surrounding a quantum system has memory of its past
evolution.  It is an important concept both because  many natural
and man-made quantum systems exist in a regime where the
assumption of a Markovian (memory-less) environment fails, but
also because it can lead to counter-intuitive results regarding
the decay of quantum effects, particularly when the quantum system
is strongly coupled to the surrounding environment. There has been
a range of efforts at constructing measures of non-Markovianity,
typically based on a scenario where the time evolution of a
quantum system is analysed for non-Markovian properties. Arguably, the most popular measures of non-Markovianity were
introduced in Refs.~\cite{Heinz09,Rivas10}. Recently, an attempt to classify
these non-Markovianity measures in a unified framework was
described in Ref.~\cite{Chrucinski14}. Useful for us here is the approach taken in ~\cite{Heinz09}, which is based on
observing the behavior of the trace distance between two quantum
states. They derived a measure of non-Markovianity by noting that all CPT maps $\Phi $ are contractions of the trace
distance  metric, and a given dynamic processs is defined as Markovian if the map is divisible,
i.e. $\Phi(\tau +t,0) = \Phi(\tau +t,t)\Phi(t,0)$, for all positive $t$ and $\tau$.
These two properties lead to the monotonicity of the trace distance, and
violations of this monotonicity indicate the
occurrence of non-Markovian dynamics.
In a similar way, below we prove that
the temporal-SW of a system undergoing a CPT map is also a
non-increasing function, i.e.,
\begin{equation}
\text{TSW}_{\rho }\geq \text{TSW}_{\Phi (\tau )\rho }
\label{theorem}
\end{equation}
for a CPT map $\Phi (\tau )$. Together with the property of divisibility,  one can conclude that the temporal-SW decreases monotonically under Markovian dynamics. Therefore our measure of non-Markovianity is defined  by
integrating the positive slope of the temporal-SW
\begin{equation}
\mathcal{N}_{\text{TSW}}\equiv \int_{\sigma _{\text{TSW}}>0}\!\!\! dt\,\sigma _{%
\text{TSW}}(t,\rho _{0},\Phi),\label{NTSW}
\end{equation}
where $\sigma _{\text{TSW}}(t,\rho
_{0},\Phi)=\frac{d}{dt}$TSW$_{\Phi(t)\rho_0}$ is the rate of
change of the temporal steerable weight. In the examples discussed
in the Supplementary Material~\cite{Supplement}, we demonstrate
explicitly how one can use this as a practical measure of strong
non-Markovianity. Here we discuss only the following example.

\emph{Proof of the monotonicity of temporal-SW under Markovian
dynamics.---} First, we prove that the temporal-SW of a system undergoing
a CPT map is a non-increasing function, as given by
Eq.~(\ref{theorem}). To obtain the temporal-SW of a qubit at time
$t_{1}$, one needs the quantity, $\sigma
_{a|x}^{\text{T}}(t_{1})-\sum_{\lambda
_{1}}D_{\lambda_1}(a|x)\tilde{\sigma}_{\lambda _{1}}$, in which
the set $\{\tilde{\sigma}_{\lambda _{1}}\}$ is chosen to maximize
Tr$(\sum_{\lambda _{1}}\tilde{\sigma}_{\lambda _{1}})$ at time
$t_{1}$. Summing all the measurement outcomes $a$ and taking the
trace, we have
\begin{eqnarray}
\text{Tr}\left[\sum_{a}\sigma
_{a|x}^{\text{T}}(t_{1})-\sum_{a}\sum_{\lambda
_{1}}D_{\lambda_1}(a|x)\tilde{\sigma}_{\lambda _{1}}\right]
\label{prove} \\
=\text{Tr}\left[\sum_{a}\sigma _{a|x}^{\text{T}}(t_{1})-\sum_{\lambda _{1}}%
\tilde{\sigma}_{\lambda _{1}}\right]=1-\mu _{1}^{\ast },  \notag
\end{eqnarray}%
where we have used the properties $\sum_{a}D_{\lambda}(a|x)=1$,
and Tr$\left[\sum_{a}\sigma _{a|x}^{\text{T}}(t_{1})\right]=1$.
Similarly, to obtain the temporal-SW of the qubit at a later time
$t_{2}=t_{1}+\tau $, one also has
\begin{equation}
\text{Tr}\left[\sum_{a}\sigma
_{a|x}^{\text{T}}(t_{2})-\sum_{a}\sum_{\lambda
_{2}}D_{\lambda_2}(a|x)\tilde{\sigma}_{\lambda _{2}}\right]=1-\mu
_{2}^{\ast },  \label{prove2}
\end{equation}%
where $\{\tilde{\sigma}_{\lambda _{2}}\}$ is chosen to maximize
Tr$(\sum_{\lambda _{2}}\tilde{\sigma}_{\lambda _{2}})$ at time
$t_{2}$. One can also perform a CPT map $\Phi (\tau
)$ to Eq.~(\ref{prove}),
giving%
\begin{eqnarray}
\text{Tr}\left[\sum_{a}\Phi (\tau )\sigma _{a|x}^{\text{T}}(t_{1})-\sum_{a}%
\sum_{\lambda _{1}}\Phi (\tau )D_{\lambda_1}(a|x)\tilde{%
\sigma}_{\lambda _{1}}\right]  \notag \\
\hspace{5mm}=\text{Tr}\left[\sum_{a}\sigma
_{a|x}^{\text{T}}(t_{2})-\sum_{a}\sum_{\lambda
_{1}}D_{\lambda_1}(a|x)(\Phi (\tau )\tilde{\sigma}_{\lambda
_{1}})\right].  \label{prove3}
\end{eqnarray}%
Since $\Phi (\tau )$ is a trace-preserving map, the value of
Eq.~(\ref{prove3}) is still $1-\mu _{1}^{\ast }$. However, we know
that the set $\{\tilde{\sigma}_{\lambda _{2}}\}$ is the optimal
way to maximize Tr$(\sum_{\lambda _{2}}\tilde{\sigma}_{\lambda
_{2}})$ at time $t_{2}$ for Eq.~(\ref{prove2}). Therefore,
comparing Eq.~(\ref{prove2}) with Eq.~(\ref{prove3}) would give
\begin{equation}
1-\mu _{1}^{\ast }\geq 1-\mu _{2}^{\ast },
\end{equation}%
This proves the theorem given in Eq.~(\ref{theorem}).
Employing the divisibility of Markovian dynamics leads to the
monotonicity of the temporal-SW:
\begin{eqnarray}
\text{TSW}\{\Phi(\tau +t,0)\sigma_{a|x}\}
&&= \text{TSW}\{\Phi(\tau +t,t)\Phi(t,0)\sigma_{a|x}\}\notag \\
&&\leq \text{TSW}\{\Phi(t,0)\sigma_{a|x}\}
\end{eqnarray}

\emph{An example of non-Markovianity of a spin-boson problem.---}
Exact solutions to the general spin-boson problem have
applications in a huge range of systems, from quantum computing to
physical chemistry and photosynthesis~\cite{Ishizaki}. Various
techniques and methods exist to numerically acquire such
solutions, one of the most powerful of which is the hierarchy
equations of motion~\cite{tanimura_pra,tanimura_jpsj}. Here we use
those equations to model a two-level system coupled to a bosonic
environment or reservoir. The general Hamiltonian is written as
\begin{equation}
H_{\text{SB}}=\frac{ E}{2} \sigma_{z}+\Delta \sigma_{x} +
\sum_{\mathbf{k}} \omega _{\mathbf{k}} a_{
\mathbf{k}}^{\dagger } a_{\mathbf{k}} +\sum_{\mathbf{k}} \sigma_{z}\otimes l_{%
\mathbf{k}}\left( a_{\mathbf{k}}^{\dagger }+%
a_{\mathbf{k}}\right),
\end{equation}
where $\Delta$ is the two-level system tunneling amplitude, and
$E$ is the two-level system splitting. The environment modes are
described with creation ($a_{\mathbf{k}}^{\dagger }$) and
annihilation operators ($a_{\mathbf{k}}$) with energy $\omega
_{\mathbf{k}}$, which couple to the system, described by the Pauli
operators $\sigma_{z}$ and $\sigma_{x}$, with strength
$l_{\mathbf{k}}$.  By assuming that the environment modes are
well-described by a Drude-Lorentz spectral density, $J(\omega
)=2\alpha \omega_c \frac{\omega }{\omega ^{2}+\omega_c^{2}}$,
where $\alpha$ is the system-reservoir coupling strength and
$\omega_c$ is the bath cut-off frequency, we can exactly solve the
dynamics of the two-level system (details can be found in
Ref.~\cite{Ishizaki,tanimura_pra,tanimura_jpsj}). We can then
compare the non-Markovianity as measured via the temporal-SW to
that given by the non-monotonic behavior of the entanglement, as
given by the concurrence~\cite{QE_Horodecki}, between the
two-level system and an isolated ancilla~\cite{Rivas10}.  One
important difference in the two approaches is that in the
temporal-SW case there is no ancilla. In the ancilla case, the
initial condition between the system and ancilla is that of a
maximally-entangled state; to mimic that in the temporal-SW case
we assume the two-level system is initially in a maximally-mixed
state.  We then evolve the entire system-reservoir equations of
motion,  using parameters relevant to energy transfer in
photosynthesis~\cite{Ishizaki}, and plot both measures in Fig.~2.

For both measures we see similar behavior, particularly as a
function of reservoir cut-off frequency and reservoir temperature.
However, as a function of system-reservoir coupling, the
entanglement measure has a larger window of detection.
\revision{This may be attributed to the hierarchical relationship between EPR steering and entanglement. For example, Ref.~\cite{Wiseman07} has shown that EPR steerable states are a superset of Bell nonlocal states, and a subset of entangled states. This hierarchy links  together these three different notions of quantum correlations. Therefore, the fact that the concurrence-based measure of non-Markovianity is more sensitive to the non-Markovianity than the temporal-SW measure seems linked, intuitively, to the notion that steering, in its EPR form, is a subset of entangled states.}
Also note
that, the sharp features in both measures are typical, and arise
because of the sudden vanishing and reappearance of both
quantities in the temporal domain.  Note that here, for
consistency with Ref.~\cite{Rivas10}, we plot $\mathcal{N}_{{\rm
TSW}}$ and $\mathcal{N}_{{\rm C}}$ using
\begin{equation}
  \mathcal{N}_{i} = \int_{t_0}^{t_f} \left| \frac{d f_i[\rho(t)]}{dt} \right| dt
+ f_i[\rho(t_f)] - f_i[\rho(t_0)],
 \label{N_i}
\end{equation}
where for the temporal-SW measure $i={\rm TSW}$, the function
$f_i[\rho(t)]$ is the temporal-SW at time $t$, while for the
concurrence measure, $i={\rm C}$, the function $f_i[\rho(t)]$ is
the concurrence between system and ancilla at time $t$.  This
definition for the integral differs from~\eq{NTSW} by a trivial
factor of $1/2$.

\begin{figure}[t]
\includegraphics[width=8cm]{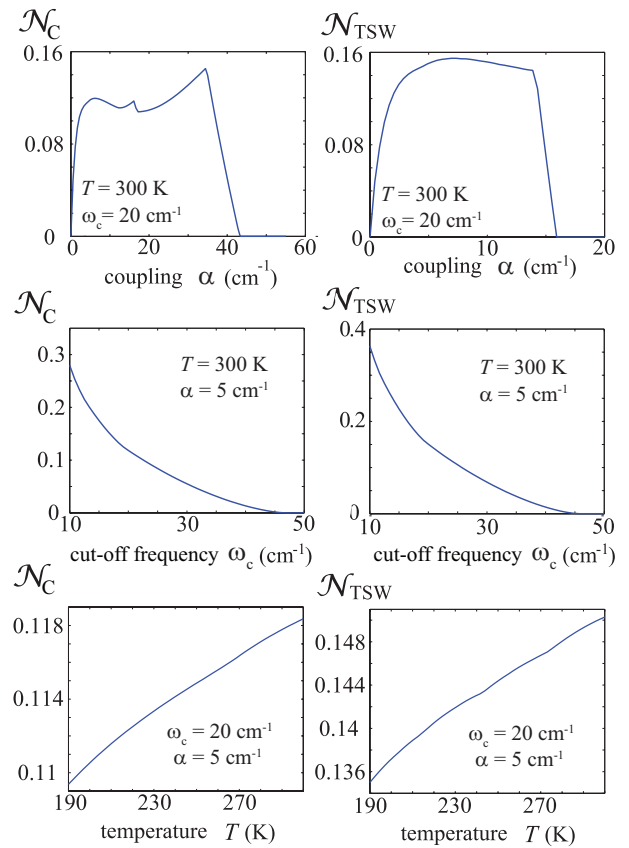}

\caption{(Color online) The non-Markovianity measures,
$\mathcal{N}_{\text{TSW}}$ (based on the temporal steerable
weight) and $\mathcal{N}_{\text{C}}$ (based on the entanglement
with an ancilla), as a function of system-reservoir coupling
$\alpha$, reservoir cut-off frequency $\omega_c$ and temperature
$T$, for a two-level system coupled to a bosonic reservoir with
Drude-Lorentz spectral density. The system parameters are chosen
to maximize the affect of the reservoir memory, with $E=0$ and
$\Delta=100$cm$^{-1}$.  The magnitude of these parameters are
typical for excitation energy transfer in
photosynthesis~\cite{Ishizaki}, where the memory effect and
structure of the environment is thought to play an important
role.}
\end{figure}

\emph{Conclusions.---} To summarize, we have discussed the
concepts of ``temporal'' steering and how this can be quantified
in a similar way to that of the original spatial EPR-steering. We
further proved that the temporal steerable weight decreases
monotonically under a CPT map and can be used as a measure of
non-Markovianity, suggesting that both forms of steering can act
as a quantum resource, similar to  entanglement. Finally we note
that, in parallel, the temporal steerable weight has been recently
implemented experimentally~\cite{Adam}.

\emph{Acknowledgement.---} The concept of a temporal steerable
weight was developed here independently of the parallel
work~\cite{Adam}. S.-L. C. thanks H.-B. Chen for the discussion on quantum non-Markovianity. This work is supported partially by the National
Center for Theoretical Sciences and Ministry of Science and
Technology, Taiwan, grant number MOST 103-2112-M-006-017-MY4. A. M.
is supported by the Polish National Science Centre under Grants
DEC-2011/03/B/ST2/01903 and DEC-2011/02/A/ST2/00305. A. M.
gratefully acknowledges a long-term fellowship from the Japan
Society for the Promotion of Science (JSPS). F. N. is partially
supported by the RIKEN iTHES Project, the MURI Center for Dynamic
Magneto-Optics via the AFOSR award number FA9550-14-1-0040, the
IMPACT program of JST, and a Grant-in-Aid for Scientific Research
(A).


%

\end{document}


\title{Quantifying Non-Markovianity with Temporal Steering: Supplementary Material}

\author{Shin-Liang Chen}
\affiliation{Department of Physics and National Center for
Theoretical Sciences, National Cheng-Kung University, Tainan 701,
Taiwan}

\author{Neill Lambert}
\affiliation{CEMS, RIKEN, 351-0198 Wako-shi, Japan}

\author{Che-Ming Li}
\affiliation{Department of Engineering Science and Supercomputing
Research Center, National Cheng-Kung University, Tainan City 701,
Taiwan}

\author{Adam Miranowicz}
\affiliation{CEMS, RIKEN, 351-0198 Wako-shi, Japan}
\affiliation{Faculty of Physics, Adam Mickiewicz University,
61-614 Pozna\'n, Poland}

\author{Yueh-Nan Chen}
\email{yuehnan@mail.ncku.edu.tw} \affiliation{Department of
Physics and National Center for Theoretical Sciences, National
Cheng-Kung University, Tainan 701, Taiwan} \affiliation{CEMS,
RIKEN, 351-0198 Wako-shi, Japan}

\author{Franco Nori}
\affiliation{CEMS, RIKEN, 351-0198 Wako-shi, Japan}
\affiliation{Department of Physics, The University of Michigan,
Ann Arbor, MI 48109-1040, USA}

\begin{abstract}
In this supplementary material we give a few illustrative examples
of the calculation of the temporal steerable weight and its
application as a measure of strong non-Markovianity for some
prototype models.
\end{abstract}

\maketitle

\section*{How to calculate the steerable weight: A~pedagogical
example}

Here we show explicitly how to calculate the steerable weight of
Skrzypczyk {\em et al.}~\cite{Paul14} in a simple example.
Specifically, we assume three types of measurements corresponding
to the projections on the eigenstates of the Pauli operators:
\begin{eqnarray}
  X &=& \ket{+}\bra{+}-\ket{-}\bra{-},
\nonumber \\
  Y &=& \ket{R}\bra{R}-\ket{L}\bra{L},
\nonumber \\
  Z &=& \ket{0}\bra{0}-\ket{1}\bra{1},
\label{pauli}
\end{eqnarray}
where $\ket{0}=\ket{H}$, $\ket{1}=\ket{V}$,
$\ket{\pm}=(\ket{0}\pm\ket{1})/\sqrt{2}$,
$\ket{R}=(\ket{0}+i\ket{1})/\sqrt{2}$, and
$\ket{L}=(\ket{0}-i\ket{1})/\sqrt{2}$, which can be interpreted
as: horizontal, vertical, diagonal, antidiagonal, right-circular,
and left-circular polarization states for the optical polarization
qubits, respectively. We can label the eigenstates of the Pauli
operators together with their eigenvalues as follows:
$\ket{x_1}=\ket{+}$ with $x_1=+1$, $\ket{x_2}=\ket{-}$ with
$x_2=-1$, $\ket{y_1}=\ket{+}$ with $y_1=+1$, \ldots, and
$\ket{z_2}=\ket{1}$ with $z_2=-1$.

Then, possible unnormalized states of Bob $\sigma_{a|x}$
($x=X,Y,Z$) for a given two-qubit state $\rho$ read
\begin{eqnarray}
  \sigma_{a|x}^{(1)} &\equiv& \sigma_{+1|X} =\tr_A[(\ket{+}\bra{+}\otimes
  I)\rho],
\nonumber \\
  \sigma_{a|x}^{(2)} &\equiv& \sigma_{-1|X} =\tr_A[(\ket{-}\bra{-}\otimes
  I)\rho],
\nonumber \\
  \sigma_{a|x}^{(3)} &\equiv& \sigma_{+1|Y} =\tr_A[(\ket{R}\bra{R}\otimes
  I)\rho],
\nonumber \\
  \sigma_{a|x}^{(4)} &\equiv& \sigma_{-1|Y} =\tr_A[(\ket{L}\bra{L}\otimes
  I)\rho],
\nonumber \\
  \sigma_{a|x}^{(5)} &\equiv& \sigma_{+1|Z} =\tr_A[(\ket{0}\bra{0}\otimes
  I)\rho],
\nonumber \\
  \sigma_{a|x}^{(6)} &\equiv& \sigma_{-1|Z} =\tr_A[(\ket{1}\bra{1}\otimes
  I)\rho],
\label{N1}
\end{eqnarray}
where $I$ is the single-qubit identity operator. A classical
random variable held by Alice,
\begin{eqnarray}
\lambda_n=[x_i,y_j,z_k]\equiv
[\bra{x_i}X\ket{x_i},\bra{y_i}Y\ket{y_i},\bra{z_i}Z\ket{z_i}],
\end{eqnarray}
can take the following values:
\begin{eqnarray}
  \lambda_1 &= [-1,-1,-1], \quad \lambda_2 &= [-1,-1,+1],
\nonumber \\
  \lambda_3 &= [-1,+1,-1], \quad \lambda_4 &= [-1,+1,+1],
\nonumber \\
  \lambda_5 &= [+1,-1,-1], \quad \lambda_6 &= [+1,-1,+1],
\nonumber \\
  \lambda_7 &= [+1,+1,-1], \quad \lambda_8 &= [+1,+1,+1].
\label{N2}
\end{eqnarray}
The extremal deterministic single-party conditional probability
distributions for Alice read\begin{eqnarray}
  [D_{\lambda_1}(+1|X),\ldots,D_{\lambda_8}(+1|X) ] &=& [0, 0, 0, 0, 1, 1, 1, 1],
\nonumber \\ {}
  [D_{\lambda_1}(-1|X),\ldots, D_{\lambda_8}(-1|X) ] &=& [1, 1, 1, 1, 0,0, 0, 0],
\nonumber \\ {}
 &\vdots&
\nonumber \\ {}
  [D_{\lambda_1}(-1|Z),\ldots, D_{\lambda_8}(-1|Z) ] &=& [1, 0, 1 ,0, 1, 0, 1, 0].\quad\quad
\label{N3}
\end{eqnarray}
Let us denote an unsteerable assemblage as\begin{eqnarray}
  \sigma_{a|x}^{{\rm US}} &\equiv&
   \sum_{\lambda}D_{\lambda}({a|x})\sigma_{\lambda} =
  \sum_{n=1}^8 D_{\lambda_{n}}({a|x})\sigma_{\lambda_n}.
\label{N5}
\end{eqnarray}
Then, we have
\begin{eqnarray}
  \sigma_{a|x}^{(1){\rm US}}  &\equiv& \sigma_{+1|X}^{{\rm US}} =
   \sigma_{\lambda_5}+ \sigma_{\lambda_6}+ \sigma_{\lambda_7}+
   \sigma_{\lambda_8},
\nonumber \\
  \sigma_{a|x}^{(2){\rm US}}  &\equiv& \sigma_{-1|X}^{{\rm US}} =
   \sigma_{\lambda_1}+ \sigma_{\lambda_2}+ \sigma_{\lambda_3}+
   \sigma_{\lambda_4},
\nonumber \\
  \sigma_{a|x}^{(3){\rm US}}  &\equiv& \sigma_{+1|Y}^{{\rm US}} =
   \sigma_{\lambda_3}+ \sigma_{\lambda_4}+ \sigma_{\lambda_7}+
   \sigma_{\lambda_8},
\nonumber \\
  \sigma_{a|x}^{(4){\rm US}}  &\equiv& \sigma_{-1|Y}^{{\rm US}} =
   \sigma_{\lambda_1}+ \sigma_{\lambda_2}+ \sigma_{\lambda_5}+
   \sigma_{\lambda_6},
\nonumber \\
  \sigma_{a|x}^{(5){\rm US}}  &\equiv& \sigma_{+1|Z}^{{\rm US}} =
   \sigma_{\lambda_2}+ \sigma_{\lambda_4}+ \sigma_{\lambda_6}+
   \sigma_{\lambda_8},
\nonumber \\
  \sigma_{a|x}^{(6){\rm US}}  &\equiv& \sigma_{-1|Z}^{{\rm US}} =
   \sigma_{\lambda_1}+ \sigma_{\lambda_3}+ \sigma_{\lambda_5}+
   \sigma_{\lambda_7}.
\label{N4}
\end{eqnarray}
The steerable weight SW can be given as the solution of the
following semidefinite program: Find
\begin{eqnarray}
{\rm SW}=1-\max \tr\Big(\sum_{n=1}^8 \sigma_{\lambda_n} \Big)
\end{eqnarray}
such that
\begin{eqnarray}
\left(\sigma_{a|x}^{(i)}-\sigma_{a|x}^{(i){\rm US}}\right)\ge 0
\quad {\rm and}\quad \sigma_{\lambda_n}\ge 0
\end{eqnarray}
for $i=1,2,\ldots,6$ and $n=1,\ldots,8$. By using a numerical
package for convex optimization~\cite{Grant08,Andersen14,Boyd04}, one can implement this
semidefinite program in a straightforward way.  This is easily
generalized to the temporal case by replacing the two-qubit
measurements in~\eq{N1} with measurements on a single qubit,
followed by evolution under the channel $\Lambda $.

\section*{Example 1: Coherent Rabi oscillations of a~Markovian system}

As a first simple example of the behavior of the temporal-SW under
a Markovian dynamics, we consider  a
qubit that undergoes coherent Rabi oscillations and purely
Markovian decay. The Hamiltonian of the system is
\begin{equation}
  H=\hbar g_1(\sigma _{+}+\sigma _{-}),
 \label{H1}
\end{equation}
where $\hbar g_1$ is the coherent coupling strength between two
eigenstates, $|+\rangle $ and $|-\rangle $, of the qubit, and
$\sigma _{+}=|+\rangle \langle -|$ and $\sigma _{-}=|-\rangle
\langle +|$ can be considered the raising and lowering operators,
respectively. A Markovian channel induces a dissipation rate
$\gamma_1$ from $ |+\rangle $ to $|-\rangle $. We assume that the
initial state, $\rho _{0}$ in Fig.~1 of the main text, is a
maximally-mixed state and then perform projective measurements
$M_{a|x}$ in three (or two) mutually-unbiased bases: $\hat{X}$,
$\hat{Y}$, and $\hat{Z}$ (or $\hat{X}$ and $\hat{Z}$). In
Fig.~1(a), we plot the temporal-SW as a function of the evolution
time $t$. We can see that the temporal-SW always remains the
maximal value of unity if there is no decay, while the temporal-SW
decreases monotonically when $\gamma_1$ is non-zero, as expected;
the dynamics of this system is Markovian.

\begin{figure}
\includegraphics[width=1\columnwidth]{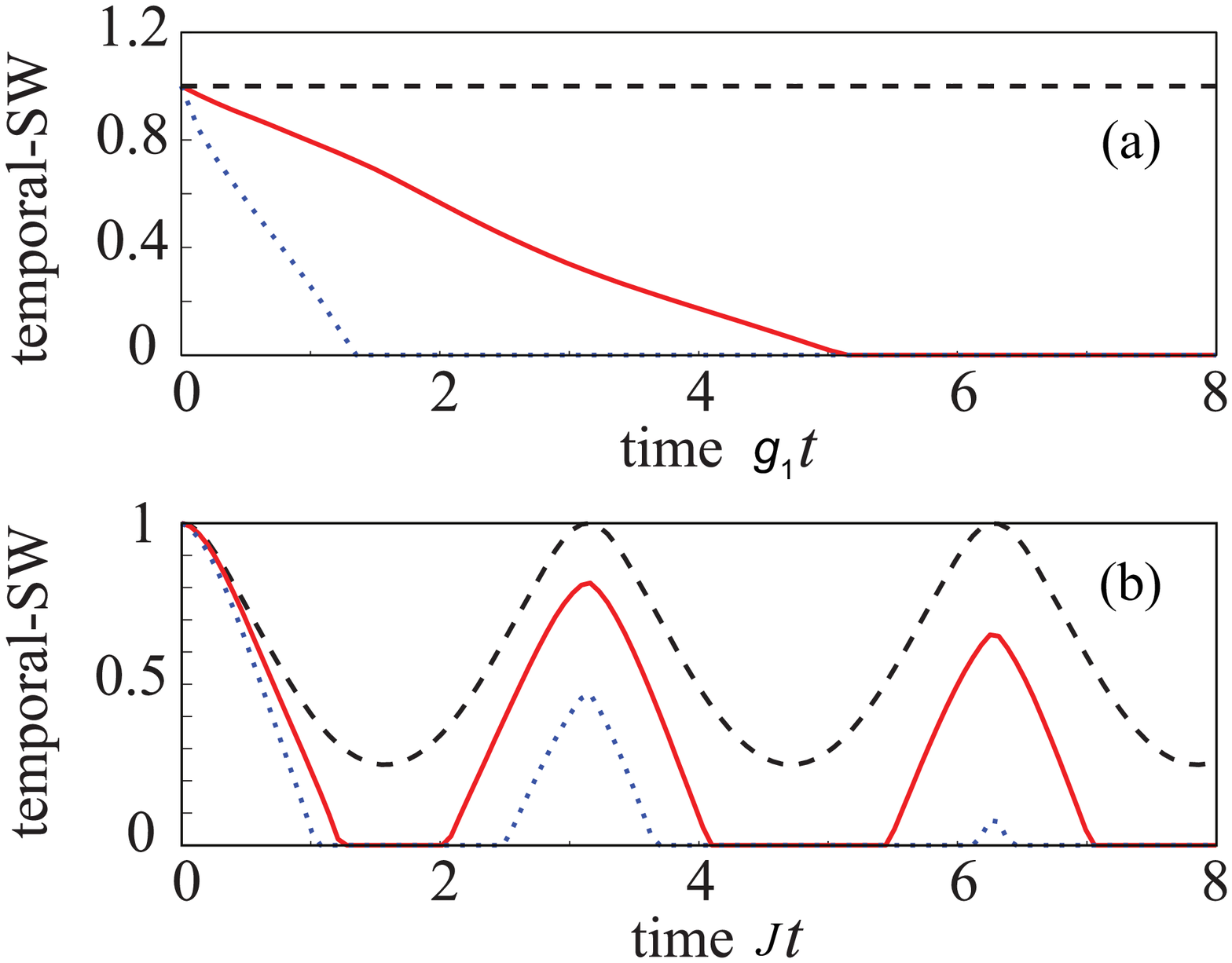}

\caption{(Color online) The temporal steerable weight
(temporal-SW) as a function of evolution time when a system is in
(a) a Markovian environment (example 1) and (b) non-Markovian
environment (example 2). (a) The temporal-SW when the system
undergoes coherent Rabi oscillations and purely Markovian decay
(example 1). The black dashed, red solid, and blue dotted curves
represent the results of the decay rate $\gamma_1/g_1 =0$, $1/6$,
and $1$, respectively. The time $t$ is in units of $1/g_1$, and
$\hbar $ is set to $1$. (b) The temporal-SW when the system
interacts with a non-Markovian environment (example 2). The black
dashed, red solid, and blue dotted curves represent the results of
the decay rate $\gamma _{2}/J=0$, $0.03$, and $0.1$, respectively.
Here, the time $t$ is in units of $1/J$.}
\end{figure}
\begin{figure}
\includegraphics[width=1\columnwidth]{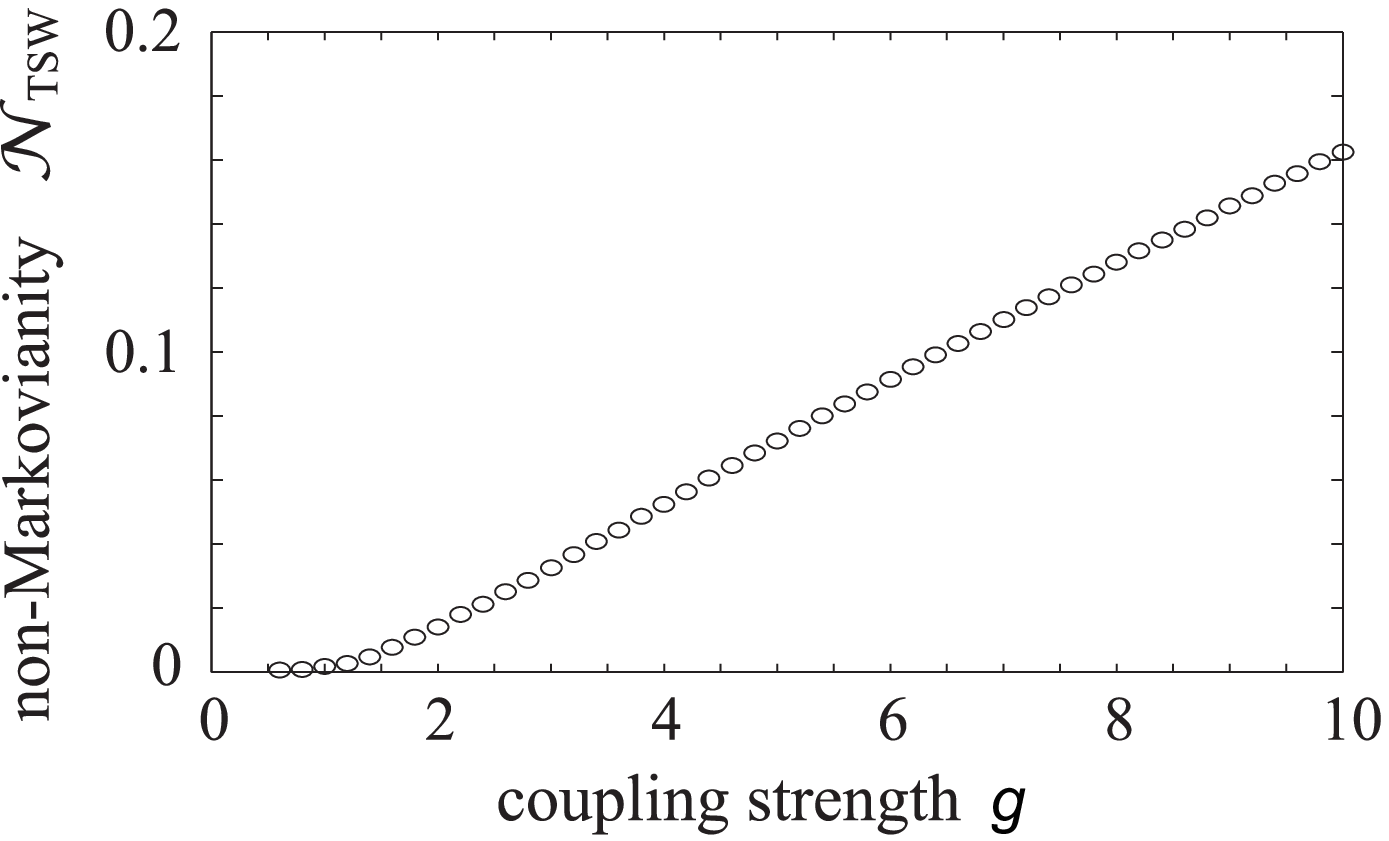}

\caption{(Color online) The degree of the non-Markovianity for a
multimode reservoir with Lorentzian spectral density (example 3).
The non-Markovianity $\mathcal{N}_{\mathrm{TSW}}$, defined by the
temporal steerable weight, as a function of the coupling strength
$g$. Here, $g$ is in units of spectral width $\omega_w$.}
\end{figure}

\section*{Example 2: A simple non-Markovian model: A~qubit coherently
coupled to another qubit}

Our second example is that of a qubit coherently-coupled to
another qubit. If we treat one qubit as the system and the other
one as the environment (by tracing it out), we have a very simple
example of a non-Markovian environment. The total Hamiltonian of
the system in the interaction picture is
\begin{equation}
H_{\text{int}}=\hbar J(\sigma _{+}^{1}\sigma _{-}^{2}+\sigma
_{-}^{1}\sigma _{+}^{2}),
 \label{H2}
\end{equation}
where $\sigma _{+}^{i}$ and $\sigma _{-}^{i}$ are the raising and
lowering operators of the $i$th qubit, and $\hbar J$ is the
coherent coupling between the system and the environment. We
assume the system qubit is also subject to an intrinsic decay with
decay rate $\gamma _{2}$. In Fig.~1(b), we plot the temporal-SW
for various decay rates $\gamma _{2}$, after tracing out the
effective environment-qubit. The initial condition of the
system-qubit is that of a maximally-mixed state, while the
environment-qubit is in its excited state. As seen in Fig.~1(b),
there is a vanishing and a reappearance of the temporal-SW of the
system qubit. Since we know that the temporal-SW should
decrease monotonically under a Markovian dynamics, the oscillation of
temporal-SW naturally shows that the qubit is undergoing
non-Markovian evolution. This memory effect in this simple example
is easy to understand in that information regarding the state of
the system-qubit flows to the environment-qubit and returns at a
later time; one cannot assume that the evolution of the
environment is not influenced by its history.

\section*{Example 3: A qubit coupled to a~non-Markovian multimode
reservoir}

In general, the dissipation $\gamma $ rate in a Master equation
description of an open-quantum system can be time-dependent, i.e.
$\gamma =\gamma (t)$. If $\gamma (t)<0$, it indicates that
information can flow back to the system and the system dynamics
can be non-Markovian. To show that the temporal-SW is sensitive to
this, we use the same example as in Breuer \emph{et
al.}~\cite{Heinz09}, where a qubit is coupled to a reservoir with
a Lorentzian spectral density. In this case, the decay rate can be
written as
\begin{equation}
\gamma (t)=-\frac{2}{G(t)}\frac{d}{dt}|G(t)|,
\end{equation}
where
\begin{equation}
G(t)=e^{-\omega_w t/2}\left[ \cosh \left( \frac{b t}{2}\right)
+\frac{\omega_w }{b}\sinh \left( \frac{b t}{2}\right) \right]
\end{equation}
with $b=\sqrt{\omega_w ^{2}-2g\omega_w }$. Here, $g$ denotes the
coupling strength and $\omega_w $ is the spectral width. We choose
a mixed state as the initial state and plot the non-Markovianity
$\mathcal{N}_{\text{TSW}}$ as a function of $g/\omega_w $ in
Fig.~2. Our results agree well with those in Ref.~\cite{Laine10}:
the non-Markovianity is zero when $g/\omega_w <0.5$, and increases
monotonically as a function of $g/\omega_w$.


%